# The Dark Matter Enigma

Jean-Pierre Luminet
*Aix-Marseille Université, CNRS, Laboratoire d'Astrophysique de Marseille (LAM) UMR 7326*
*& Centre de Physique Théorique de Marseille (CPT) UMR 7332*
*& Observatoire de Paris (LUTH) UMR 8102*
*France*
*E-mail:* jean-pierre.luminet@lam.fr

**Abstract**

In this pedestrian approach I give my personal point of view on the various problems posed by dark matter in the universe. After a brief historical overview I discuss the various solutions stemming from high energy particle physics, and the current status of experimental research on candidate particles (WIMPS). In the absence of direct evidence, the theories can still be evaluated by comparing their implications for the formation of galaxies, clusters and superclusters of galaxies against astronomical observations. I conclude briefly with the attempts to circumvent the dark matter problem by modifying the laws of gravity.

## A cosmological conundrum

The existence of dark matter was first suspected in the early 1930s. While measuring the velocities of galaxies grouped together in clusters, the Swiss astronomer Fritz Zwicky deduced the presence of an invisible mass. His calculations showed that if only the visible mass of the galaxies was taken into account, they should have separated due to their high proper velocities [1]. Yet the galaxies remained grouped together and the clusters were stable. The most plausible explanation was the presence of an invisible mass that gravitationally bound the galaxies but that did not emit electromagnetic radiation. Subsequent measurements have shown that there must be at least ten times more dark matter than luminous matter in galaxy clusters [2].

Further indirect evidence for the existence of dark matter came from the rotational velocities of spiral galaxies. These systems usually rotate at higher velocities near their centers than at their edges. In the late 1960s, Vera Rubin, W. Kent Ford, and Norbert Thonnard found that the rotational velocity remains more or less constant beyond a certain distance from the center, even in far flung regions containing few stars [3]. This implied that dark matter was present well beyond the limits of the galactic disc, perhaps forming an extended halo.

The notion of dark matter has since become ubiquitous in astronomy and cosmology. Dark matter plays an important role in galaxy formation models, its presence ensuring that the first structures condensed over the short timeframes reflected in observations [4]. From studying gravitational mirages, physicists can dispatch the distribution of visible and invisible matter; and they have concluded that at least ten times more dark matter than luminous matter is needed to account for observations [5]. A significant proportion of dark matter is also needed in the initial composition of the universe to explain the anisotropy spectrum of the cosmic diffuse background — that is, the distribution of temperature irregularities in the first radiation emitted by the universe when it was only 380,000 years old [6].

Recent surveys make possible a precise estimation of the composition of the observable universe [7]. Luminous matter represents only 0.5% of the total; dark energy, a form of non-material and antigravitational energy, 68%. The remaining 32% is attributed to dark matter—a contribution that is almost entirely invisible to telescopes.

The true nature of dark matter is one of the most significant challenges in modern astrophysics, especially since its distribution is far from uniform. Consider the galaxies Dragonfly 44 and NGC1052-DF2. Discovered in 2015, Dragonfly 44 has a mass comparable to the Milky Way, but appears to be almost exclusively dark matter (99.99%) [8]. By contrast, NGC1052-DF2, which was discovered in March 2018, appears to be devoid of dark matter [9].

It might seem reasonable to imagine that dark matter is composed from all the celestial bodies too faint to be observed. Interstellar dust would be included, along with clouds of cold molecular hydrogen, and the dark stars in the halos of galaxies known as MACHOS (Massive Compact Halo Objects). This grouping includes black dwarfs, neutron stars, black holes, and brown dwarfs—stars too small to emit a glow from their nuclear reactions. Observations of the Large Magellanic Cloud have found that MACHOS amount to less than 8% of the black mass of the galactic halo. The projects involved include EROS (Experiment in Search of Dark Objects) and OGLE (Optical Gravitational Lensing Experiment). Above one thousandth of a solar mass, MACHOS are massive enough to act as gravitational microlenses. This effect can be used to obtain an upper bound to their presence in the halo of galaxies [10].

These celestial bodies are composed of protons and neutrons, collectively known as baryons. Big bang cosmological models have successfully explained the observed abundance of light chemical elements—deuterium, helium and lithium—that were produced in a brief primordial nucleosynthesis. The proportions of these elements can be used to determine the density of baryonic matter in the young cosmos. A value for its current density, both visible and black, can be determined using Hubble-Lemaître's law. These results agree with observations. Primordial nucleosynthesis imposes an upper limit on the current density of baryonic matter: it cannot exceed 5% [11].

If 32% of the universe is dark matter and baryonic matter is only 5%, that leaves 27% unaccounted for. There are two ways to solve this problem. The first is to abandon the very idea of dark matter—astronomers must have made a mistake and Newton's law of universal attraction requires modification. A second approach originates in particle physics and involves the introduction of non-baryonic dark matter.

# A Solution from Particle Physics

Recent observational evidence has shown that, aside from gravitational force, dark matter interacts only barely, if at all, with ordinary matter. This suggests that dark matter is predominantly non-baryonic in nature, even if it produces gravitational effects similar to those associated with mass.

The theoretical candidates for non-baryonic dark matter fall into three classes.

• Hot Dark Matter (HDM) does not refer to a specific temperature, but to low-mass particles with velocities close to the speed of light. Obvious candidates are the three types of neutrinos in the standard model. Neutrinos have been detected experimentally, even though their limited interactions with the electromagnetic field renders them elusive. On average, there are 400 neutrinos per cubic centimeter of space throughout the universe. Neutrinos would only need a tiny mass to account for all dark matter [12]. Indeed the first cosmology article I published in 1981 was precisely about the possibility that neutrinos, about which little was known yet, would have sufficient mass to close the universe, i.e. give it a density higher than the critical density [13]. Experiments to measure neutrino mass began in 1988 at the Super-Kamiokande detector in Japan and continue today at the Karlsruhe Tritium Neutrino Experiment (KATRIN) in Germany. These studies have yielded a series of gradually decreasing mass values [14]. Neutrino mass is now thought not to exceed $10^{-34}$ grams—a billionth the mass of a proton. Although negligible, the contribution of neutrinos to the energy density of the universe is on the same order as that of light matter (0.5%).

• Warm Dark Matter (WDM) is composed of a hypothetical particle, the sterile neutrino [15]. In contrast to the neutrinos of the standard model, sterile neutrinos do not interact with the weak and strong electromagnetic interactions, making them even more difficult to detect. Due to their mass, sterile neutrinos interact via gravity, meaning that, given a sufficient mass, they could account for dark matter. The researcher who provided the first tools to understand them was the Italian physicist Ettore Majorana, one of the great figures in twentieth century physics. In 1938, Majorana disappeared under mysterious circumstances when he was not yet 32 years old. Like Isaac Newton, he preferred not to make his ideas public and the importance of Majorana's work was not properly understood for decades; for a brief biography of Ettore Majorana, see [16]. The existence of sterile neutrinos is due to an anomaly in the helicity of ordinary neutrinos. Particles in the standard model have an intrinsic kinetic moment termed spin—as if they were tiny rotors with an axis of rotation parallel to their velocity vector. A particle's helicity is said to be either left or right-handed depending on the direction of rotation. Although the other particles in the standard model—leptons and quarks—can be both left and right-handed, no right-handed neutrinos have ever been observed.

The minimum standard model predicted incorrectly that ordinary neutrinos would have a strictly zero rest mass. In 1998, the discovery of their oscillations forced physicists to assign neutrinos mass, after all. Sterile neutrinos were introduced to account for these masses and oscillations, following on from hypotheses previously made by Majorana [17]. Models that make use of sterile neutrinos have several advantages. Their presence helps explain, for example, the predominance of matter over antimatter, a process known as baryogenesis. If the mass of sterile neutrinos is found to be greater than $10^{-29}$ grams, they would be potential dark matter particles. Several experiments have sought to detect sterile neutrinos, but their results are contradictory. In 2016, researchers at the IceCube Neutrino Observatory in Antarctica found no evidence for the existence of sterile neutrinos. Yet two years later, researchers at the MiniBooNE collaboration found a much larger than expected oscillation for neutrinos, an indication in their favour [18].

• Cold Dark Matter (CDM) is composed of larger and slower moving particles than HDM or WDM. These exotic particles only appear in high-energy physics theories beyond the standard model. The two main candidates are axions and neutralinos. The former are hypothetical particles assumed to be stable, electrically neutral, and extremely light, with a mass between $10^{-39}$ and $10^{-36}$ grams. They were introduced by theorists to help explain charge-parity symmetry (CP) in strong interactions. Frank Wilczek adopted the name Axion from a brand of detergent to reflect that the problem plaguing quantum chromodynamics (the theory of strong interaction), namely the violation of CP symmetry, had been resolved [19]. It has been theorized that the Big Bang created sufficient axions to account for the missing dark matter [20]. Experiments such as CAST and ADMX have been unable to detect axions, which only interact weakly with other matter [21].

The existence of the neutralino is predicted by the theory of supersymmetry (SUSY), an attempt to unify the three fundamental interactions of the standard model [22]. Under SUSY, all known fundamental particles have heavier supersymmetric counterparts, known as superpartners or sparticles. The theory has considerable appeal for researchers because it may offer solutions to several enduring problems, such as the disparity in the calculated mass of the Higgs boson under quantum field theory. SUSY could also account for non-baryonic dark matter using the neutralino [23], an electrically neutral combination of the photino (superpartner of the photon), the zino ($Z_0$ boson), and the higgsino (Higgs boson). Neutralinos only form at very high energy levels, such as those found during the Big Bang. As a result of their assumed stability, neutralinos should be abundant; and with a mass a hundred times greater than protons ($10^{-22}$ grams), the neutralino may be the missing dark matter particle.

## The Search for WIMPS

Neutralinos can be detected either directly, from interactions in a detector and collisions in a particle accelerator, or indirectly, by means of the particles created by their disintegration or annihilation—photons, neutrinos, positrons, and antiprotons.

The insensitivity of neutralinos to electromagnetic interaction hampers any attempt at direct detection. Neutralinos belong to a vast family of particles known as WIMPS (Weakly Interacting Massive Particles). Despite the extremely low probability of their interaction, it should be possible, with a large enough detector and considerable patience, to observe a WIMP come in contact with a particle of baryonic matter. A number of experiments are currently underway. These experiments include EDELWEISS in France, XENON in Italy, LUX in the US, and PandaX in China. Nothing has been found so far.

Another solution to the problem of detecting WIMPS involves producing them artificially from the collisions of particle beams in accelerators. A funny anectdote is the following. In 1994, Kenneth Lane and David Gross made a bet on SUSY during a dinner at the summer school of theoretical physics in Erice, Sicily. At the time, the predicted sparticle had not been detected, but plans for the LHC had been drawn up. Lane proposed that if the new collider proved the theory, he would pay for a meal at Girardet, a restaurant in Lausanne widely considered among the world's best. If the theory was not proven, Gross would pick up the check. Written on a tablecloth, the terms of the bet stipulated that it would be payable as soon as the LHC produced sufficent high-energy collisions to be certain of the outcome. The minimum number was estimated at $5.10^{24}$. A quarter of a century later, Girardet no longer exists. Since 2010, the LHC has collected more than $15.10^{24}$ pieces of data without finding any trace of a spark, even when the energy range should have allowed for it. Lane claims that the bet is now payable, if not at Girardet's, then at another restaurant of a similar calibre. But Gross is not about to concede. The data has been gathered, but the analysis is not yet complete (the amount of data produced by a particle accelerator is measured in a dark unit known as a reverse femtobarn, that corresponds to about $10^{23}$ collisions).

Attempts to detect WIMPS in this manner are the focus of ongoing research at CERN's Large Hadron Collider (LHC) [24]. To date, researchers at the LHC have not found anything either. Despite these disappointing results, SUSY can never really be completely overturned. The theory contains numerous free parameters and can be adjusted so that its predicted particles only appear at energy levels beyond the reach of current colliders. Such arbitrary adjustments inevitably detract from the elegance of the theory, compromising the attribute that made it attractive in the first place.

## Antimatter

Another potential indirect means of detecting WIMPS involves antimatter. In theory, each particle has an associated antiparticle with the same mass but the opposite electrical charge: the positron for electrons, antiquark for quarks, antiproton for protons, and so on. Antimatter is composed of these antiparticles. Antihydrogen, for example, has a positron in orbit around an antiproton. When antimatter comes into contact with matter, the two annihilate each other, creating radiative energy. When mass is created from energy, as occurs in particle accelerators, it is equally distributed between particle pairs and antiparticles. Antihydrogen was produced for the first time at CERN in 1996. The Alpha collaboration at CERN has recently managed to confine a thousand antihydrogens for several hours by using a magnetic trap.

Following the Big Bang, the universe was in a hot plasma state, behaving like a vast natural particle accelerator. Although it should have generated just as much matter as antimatter, the composition of the current universe suggests otherwise. On average, there seems to be only one antiparticle for every billion particles. As a result, there would be no stars or galaxies made of antimatter [25]. High-energy physics models have been developed to explain this asymmetry. As none of the models are completely convincing, some authors have attempted in various ways to attribute a repulsive gravitational mass to antimatter [26]. Although it seems unlikely, this hypothesis has the advantage of being easy to test in the laboratory. CERN's Alpha experiment will attempt to find out what happens when artificially produced antiatoms are dropped in a gravitational field [27]. A result that contradicts standard theories would be truly revolutionary, but the chances seem remote. The most likely finding is that antimatter does not contribute at all to dark matter density.

Despite the lack of results, dark matter researchers are keenly interested in the positron and antiproton fluxes observed in some cosmic rays. Both antiparticles are produced by conventional astrophysical sources such as pulsars, but they might also emerge from the disintegration of exotic dark matter. In 2011, the AMS (Alpha Magnetic Spectrometer) instrument was installed on the International Space Station (ISS). To date, the AMS has detected several billion antiparticles, but no anomaly in the flow of antiparticles has revealed their origin [28].

## The Candidates

Three categories of theories have been developed in an effort explain non-baryonic dark matter: Cold Dark Matter, Hot Dark Matter, and Warm Dark Matter. In the absence of direct experimental evidence, these theories can still be evaluated by comparing their implications for the formation of galaxies, clusters and super clusters of galaxies against astronomical observations.

In the case of HDM, the contribution of ordinary neutrinos to energy density is negligible given their low mass energy. They cannot account for non-baryonic dark matter. There may well be other WIMPS, but their numbers are constrained by the formation of galaxies [29]. High speed particles slow the formation process and fragment agglomerates of matter. Simulations have shown that if hot dark matter was prevalent, galaxy superclusters would have formed early in the history of the universe, only to break apart into smaller clusters and galaxies [30]. Such a scenario conflicts with observations from both ground and space-based telescopes, which indicate that first stars, then galaxies, must have formed less than a billion years after the big bang [31]. HDM will likely be ruled out as a candidate for dark matter.

According to CDM theories, cold WIMPS emerged from the big bang at speeds significantly slower than that of light. These particles aggregated into galactic masses faster than hot matter, meaning that galaxies formed before clusters—a scenario supported by observations. Numerical simulations involving CDM also provide a credible account of supergalactic-scale structuring in the early universe. CDM has long seemed the most plausible of the three candidates.

None of these particles have ever been detected.

Both WDM and CDM theories correctly explain the formation of structures on a supergalactic scale. One cannot be chosen over the other on this basis alone [32]. Another approach involves numerical simulations on a subgalactic scale, which can then be compared to observational counts of dwarf galaxies in orbit around large galaxies. These comparisons have also proven inconclusive. Some simulations are indeed compatible with CDM assumptions, but only a portion of WDM models are eliminated. How then to distinguish between CDM and the remaining WDM models? Three scenarios are currently being investigated: the formation of dark halos, or dark matter halos, that envelop galactic discs and extend well beyond the visible limits of galaxies [33]; gravitational shear, the distortion of images of distant galaxies by foreground mass concentrations [34]; and tidal stellar streams, groupings of stars orbiting a galaxy born from an ancient dwarf galaxy predecessor and elongated along its orbit by tidal forces [35].

Although it is not currently possible to distinguish between CDM and WDM models on scales greater than a few million light years, the pair offer different predictions on the scale of gravitational microlenses. If it turns out that future results eliminate pure CDM theory, the remaining explanations for non-baryonic dark matter will be limited. Sterile neutrinos will remain a possible constituent, along with a mixture of different forms of dark matter, provided, of course, that none of the ingredients are mutually exclusive.

## Modified Gravity

Faced with so many unresolved questions and problems, researchers have challenged the very notion of dark matter, considering, instead, a change to the laws of gravity. First proposed by Mordehai Milgrom in 1983, Modified Newtonian Dynamics (MOND) assumes that Newton's second law must be corrected at low accelerations—beneath a threshold several orders of magnitude below earth's gravity [36]. Such a regime would be in effect in the outmost regions of spiral galaxies. Modified gravity may also explain the stellar velocity profile measured by Rubin, Ford, and Thonnard but without recourse to dark matter.

MOND has been modified many times in response to conflicting observations. Although the theory works well for galaxies, it is less effective at larger scales. Under MOND, the dynamics of galaxy clusters cannot be reproduced without incorporating dark matter. This is also true for explanations of the anisotropy spectrum found in the cosmic diffuse background, gravitational lensing, and the formation of large structures [37].

As observations become ever more refined, MOND is becoming ever more unstable. The absence of dark matter in NGC1052-DF2, for example, contradicts the theory because its gravitational signature should be present in all galaxies. The same is true of super-spiral galaxies much larger than our Milky Way. Astronomers recently measured the rotation speed of 23 super-spirals and found very fast rotations involving the presence of a large quantity of matter [38].

Despite these issues, MOND has not been discarded. Based on the observation that MOND works best on the scale of galaxies and dark matter on larger scales, various approaches have been proposed to reconcile the two competing models. In 2014, Justin Khoury proposed a new theory of superfluid dark matter [39]. Superfluids are liquids that exhibit zero viscosity when cooled to relatively low temperatures, such as 2 K (–271°C) for helium-4. According to Khoury, dark matter is superfluid at the scale of galaxies, but too hot to maintain these properties at larger scales, whereupon it reverts to conventional dark matter. The idea of a superfluid dark matter had been proposed previously, but Khoury's model was able to reproduce MOND's predictions in galaxies without any need for modified gravity.

At the beginning of the twenty-first century there was much talk of an apparent acceleration anomaly affecting the Pioneer 10 and 11 space probes during their transit through the transneptunian solar system. The anomaly, in this case a very slight deceleration, was measured between 1979 and 2002 and gave rise to numerous studies and speculation. In particular, it was a perfect opportunity for supporters of MOND theory to see it as evidence of the modification of Newton's law in weak gravitational fields. In 2011 the mystery was solved by several teams: the origin of the observed deviation was due to the pressure of infrared radiation emitted by the radioisotope thermoelectric generator (RTG), the small nuclear electric generator embedded on the probes to free itself from solar panels that had become inactive at a long distance from the Sun. The anomaly had nothing to do with modified gravity or dark matter! [40]

Another way to avoid the use of hypothetical dark matter is to view gravity not as a fundamental interaction, but as an emerging phenomenon of fundamental quantum information bits encoded in the intimate structure of spacetime. In the recent entropic gravity models of Erik Verlinde, dark matter is considered an illusion arising from the dynamics that link dark energy and ordinary baryonic matter [41]. Although not yet fully developed, this theory has already successfully reproduced the rotation curves of spiral galaxies [42]. The next step is to construct a theory capable of describing the evolution of the primordial universe.

*This article is an adapted version of the original paper published in Inference - The International Review of Science Vol. 5 n°3 (September 2020).*